\documentclass[twocolumn]{aa}
\usepackage{graphics}

\newcommand{\etal}{et al.}

\newcommand{\ltsim}{\raisebox{-1mm}{$\stackrel{<}{\sim}$}}

\newcommand{\eg}{e.g.}
\newcommand{\ie}{i.e.}
\newcommand{\etc}{etc.}
\def\arcm{\hbox{$^\prime$}}
\def\arcs{\arcm\hskip -0.1em\arcm}

\begin{document}

\title{The XMM-Newton EPIC Background: Production of Background Maps
and Event Files \thanks{Based on observations with XMM-{\it Newton},
an ESA Science Mission with instruments and contributions directly
funded by ESA Member States and the USA (NASA).}}

\author{A.M. Read\inst{1,2} and T.J. Ponman\inst{1} }

\institute{$^1$School of Physics and Astronomy, University of
Birmingham, Edgbaston, Birmingham B15 2TT, UK \\ $^2$Deptartment of
Physics and Astronomy, Leicester University, Leicester LE1 7RH, UK}
     
\offprints{A.M.\,Read}
\mail{amr30@star.le.ac.uk}

\date{Received; accepted }
\titlerunning{The XMM-Newton EPIC Background}

\abstract {

 XMM-Newton background maps for the 3 EPIC instruments in their
 different instrument/mode/filter combinations and in several energy
 bands have been constructed using a superposition of 72 pointed
 observations. Event datasets, with point sources excised, for the
 different instrument/mode/filter combinations have also been
 assembled, with longer exposure times than previously available
 files. The construction of the background maps and event files,
 together with their properties and usage are described here. Also
 given are statistical properties of the photon and particle
 components of the XMM-Newton EPIC background, based on the analysis
 of the 72 datasets.

          } 

\maketitle 

\section{Introduction} 
\label{intro}

The XMM-Newton observatory (Jansen \etal\ 2001) provides unrivalled
capabilities for detecting low surface brightness emission features
from extended and diffuse galactic and extragalactic sources, by
virtue of the large field of view of the X-ray telescopes with the
EPIC MOS (Turner \etal\ 2001) and pn (Str\"{u}der \etal\ 2001) cameras
at the foci, and the high throughput yielded by the heavily nested
telescope mirrors. The satellite has the largest collecting area of
any imaging X-ray telescope.

In order to exploit the excellent EPIC data from extended objects, the
EPIC background, known now to be higher than estimated pre-launch,
needs to be understood thoroughly. With a good model of the particle
and photon background, one can correctly background-subtract images
and spectra extracted over different energy bands and from different
areas of the detectors.

Here we provide details of a project to use a large number of
XMM-Newton pointed observations to help define the EPIC background and
to produce background maps for each of the three EPIC instruments (pn,
MOS1 \& MOS2) in several different energy bands. Also, significantly
improved background event files, useful for spectral analysis, with
longer exposures than previously produced files, and specific to
several particular instrument/mode/filter combinations, have been
created as part of the analysis. The co-addition of many fields allows
the minimization of any `cosmic variance', resulting from variations
in the local diffuse X-ray emission or contamination from
pathologically bright sources.

This paper is intended as an aid when studying or working with the
EPIC background, and in particular, when using these files. The
current understanding of the XMM-Newton background is described
briefly in Section 2. Section 3 describes the methods used in the
creation of the background products. Information on how to use the
background images for XMM EPIC background analysis is given in Section
4, along with caveats as to their use. The relevant information for
the event files can be found in Section 5, followed by some concluding
remarks.

All the background product files (maps, event files, related software
\etc), together with other scripts and procedures for XMM-Newton
Background Analysis are available from
\verb£http://www.sr.bham.ac.uk/xmm3/£

\section{The XMM-Newton EPIC X-ray Background} 

The EPIC background has been shown (via the work of Lumb \etal\ (2002)
and many others; see Appendix) to comprise solar soft protons,
cosmic rays, electronic noise and cosmic X-ray photon background. We
now briefly discuss the properties of these contributions:

\begin{itemize}

\item Solar soft protons (see Sect.3.2) perhaps accelerated by
`magnetospheric reconnection' events and trapped by the Earth's
magnetosphere, then gathered by XMM-Newton's grazing mirrors. These
dominate the times of high background. 

\end{itemize}

During quiescent periods (\ie\
with no significant soft proton contamination), the remaining
components are:

\begin{itemize}

\item High energy, non-vignetted cosmic ray induced events, unrejected
by the the on-board electronics. Also, associated instrumental
fluorescence, due to the interaction of high-energy particles with the
detector.

\item Non-vignetted electronic noise: bad (bright) pixels, and dark
current, though the latter is thought negligible. In actuality,
most of the bright pixels are rejected on board the satellite, and the
vast majority of the remaining bad pixels are removed by analysis
software on the ground. The amount of `flickering' activity of these
remaining pixels is thought to be related to the radiation level of
the observation.
 
\item Low to medium energy, vignetted X-ray photons from the
sky. These can be divided into the local (predominately soft) X-ray
background, the cosmic (harder) X-ray background, and single
reflections entering the telecope from bright sources outside of the
nominal field of view (FOV). Lumb \etal\ (2002) estimate that the
contribution of diffuse flux gathered from out-of-field angles of
0.4$-$1.4 degrees is of order 7\% of the true, focussed in-field
signal, and the associated systematic error (due mainly to the energy
dependence) is $\pm$2\%.

\end{itemize}

Table~1 gives a brief summary of the temporal, spatial and spectral
properties of these, the major components contributing to the
XMM-Newton EPIC background.

\begin{table*}
\caption[]{Summary of the components within the XMM-Newton EPIC
Background; temporal, spatial and spectral properties}
\begin{tabular}{|l|l|l|l|l|l|}
\hline
  & \multicolumn{2}{c|}{PARTICLES} &  & \multicolumn{2}{c|}{PHOTONS} \\ 
\hline
                & SOFT PROTONS        & INTERNAL              & ELECTRONIC                 & HARD X-RAYS       & SOFT X-RAYS \\
                &                     & (Cosmic-ray induced)  & NOISE                      &                   &             \\ \hline
\noalign{\smallskip}
Source          & Few 100\,keV          & Interaction of High & 1) Bright pixels           & X-ray       & Local Bubble \\ 
                & solar protons         & Energy particles    & 2) Elec.\,overshoot        &  background & Galactic Disk \\ 
                &                       & with detector       & near pn readout            &  (AGN etc)  & Galactic Halo \\ \hline
\noalign{\smallskip}
Variable?        & & & & & \\ 
(per Obs)       & Flares ($>$1000\%)    & $\pm$10\%           & $\pm$10\%                  & Constant            & Constant \\
(Obs to         & Unpredictable.     & $\pm$10\%           & 1) $>$1000\% (pixels       & Constant            & Variation with \\
\verb£   £Obs)  & More far from      & No increase after   & come and go, also &                              & RA/Dec ($\pm$35\%) \\ 
                & apogee.            & solar flares        & meteor damage) & & \\ 
                & Low-E flares turn  & & & & \\ 
                & on before high-E   & & & & \\ \hline
\noalign{\smallskip}
Spatial         & & & & & \\ 
Vignetted?      & Yes (scattered)               & No                    & No         & Yes           & Yes\\ 
Structure?      & Perhaps,              & Yes. Detector +       & Yes & No                   & No, apart from \\  
                & unpredictable         & construction &       1) Individual pixels            &     & real astronomical \\  
                &                       & MOS: outer CCDs more        & \& columns        &        & objects \\  
                &                       & Al, CCD edges more Si         & 2) Near pn     & & \\  
                &                       & PN: Central hole in & readout (CAMEX) &    &  \\ 
                &                       & high-E lines ($\sim$8\,keV) &   &  & \\    \hline
\noalign{\smallskip}
Spectral        & Variable              & Flat + fluorescence +        & 1) low-E ($<$300\,eV), & $\sim$1.4 power law.& Thermal with \\  
                & Unpredictable         &  detector noise              & tail may reach & Below 5keV,        &  \ltsim 1keV emission \\  
                & No correlation         & MOS: 1.5\,keV Al-K   & higher-E & dominates over      & lines \\  
                & between intensity            &     1.7\,keV Si-K    & 2) low-E ($<$300\,eV) & internal   & \\  
                & + shape     &     det.noise$<$0.5\,keV.    & & component & \\  
                & Low-E flares turn      &     High-E $-$ low-intensity & & & \\  
                & on before high-E      &     lines (Cr, Mn, Fe-K, Au) & & & \\  
                &                       & PN:  1.5\,keV Al-K           & & & \\  
                &                       &     no Si (self-absorbed)    & & & \\  
                &                       &     Cu-Ni-Zn-K ($\sim$8\,keV)& & & \\  
                &                       &     det.noise$<$0.3\,keV     & & & \\  
\hline
\end{tabular}
\end{table*}

The present analysis is related in several respects to the work of
Lumb \etal\ (2002), and the reader is encouraged to consult this work.
Additional notes regarding other related work on the EPIC background
can be found in the Appendix.

\section{Data Analysis} 

Source-subtracted, high-background-screened and exposure- and
source-corrected images (maps) of the particle and photon components
of the EPIC background have been created separately for each EPIC
instrument, and in several different energy bands. This has been
performed separately per observation, over a large number of
individual observations.

The individual background maps for a particular instrument and in a
particular energy band have then been collected together (for the same
instrument mode/filter combination) over the whole set of
observations. Via various cleaning, filtering and `$\sigma$-clipping'
techniques, a `mean' background map is created (for each particular
background component/instrument/energy band/mode/filter combination).

Procedures have been written to perform the different aspects of the
analysis, making extensive use of the XMM-SAS tasks, Chandra CIAO
tools and HEASOFT's FTOOLS utilities.

Before discussing the analysis in depth, we describe some of the files
and terms used, and the structure of some of the final products.

\subsection{Data Analysis: Files \& Definitions}

We first define a number of terms used below:

\begin{itemize} 

\item Background maps. Detector maps of all the background components
combined \ie\ soft and hard X-ray photons (vignetted), internal
particles (non-vignetted), some small component of soft protons
(scattered/funnelled) and single-reflections from out-of-FOV sources.

\item Detector maps. Images in detector coordinates (DETX DETY) (as
opposed to sky coordinates), defined, in the present analysis, such
that the individual pixels are 1\arcm\ by 1\arcm, DETX/DETY of (0,0)
lies at a pixel intersection, and the full area of the CCDs is
covered. A scale of 1 arcminute has been chosen so as to give us
decent count statistics in each pixel over the full energy band. More
detailed information as to the coordinate system used (and the 4\arcs\
detector maps used within the analysis) is given in Table~2. Software
has been written (see Sect.~4.2) to rebin the detector maps to any
spatial scale, and to reproject the maps to any sky coordinates
(specified via a user-input sky image).

\begin{table}
\caption[]{The coordinate systems of the 1\arcm\ background detector
maps and the 4\arcs\ detector maps used in the analysis. }
\begin{tabular}{cccccc}
\hline
Instr. & Coord. & \multicolumn{2}{c}{Range in        } & \multicolumn{2}{c}{No.\,of pixels} \\
           &(DETX/Y)& \multicolumn{2}{c}{Detector coords.} & \multicolumn{2}{c}{in maps} \\ 
       &        & min. & max. &(4\arcs)  & (1\arcm)  \\ \hline

PN & DETX & -19199 & 14400 & 420 & 28 \\
   & DETY & -16799 & 15600 & 405 & 27 \\
M1 & DETX & -20399 & 20400 & 510 & 34 \\
   & DETY & -20399 & 20400 & 510 & 34 \\
M2 & DETX & -20399 & 20400 & 510 & 34 \\
   & DETY & -20399 & 20400 & 510 & 34 \\ \hline
\end{tabular}
\end{table}

\item Instruments. The present analysis has been performed for each of
the EPIC instruments; pn, MOS1 \& MOS2 (hereafter PN, M1, M2).

\item Instrument mode \& filter. Several different data acquisition
modes exist for each of the EPIC instruments. Also, observations have
been taken using different filters for each of the EPIC
instruments. The present analysis has been performed for the most
common instrument mode/filter combinations, and those most useful to
the analysis of diffuse X-ray emission from extended objects. These
are full-frame mode with thin filter (hereafter ft) and full-frame mode with
medium filter (fm), for each of PN, M1 \& M2, and also (for PN)
full-frame-extended mode with thin filter (et) and full-frame-extended
mode with medium filter (em).

\item Energy bands. The analysis has been performed in the following
XMM-XID \& PPS standard energy bands; energy bands 1$-$5, defined as
follows; 1: 200$-$500\,eV 2: 500$-$2000\,eV 3: 2000$-$4500\,eV 4:
4500$-$7500\,eV 5: 7500$-$12000\,eV, and a full energy band 0:
200$-$12000\,eV.

\item Closed datasets. A few XMM observations have been obtained with
the filter wheel in the `closed' position. No photons reach the CCDs,
so the event files contain only the instrumental and particle
components of the background. Such datasets have been collected
together and processed, and exist on
\verb£ftp://www-station.ias.u-psud.fr/pub/epic/£
\verb£Closed£ (Marty \etal\
2002). They are dependent on the instrumental mode, and have been
processed for M1 full-frame (exposure 110655\,s), M2 full-frame
(101336\,s), PN full-frame (42079\,s) and PN full-frame-extended mode
(26731\,s).

\end{itemize}

\subsection{Data Analysis: Preparation and Reduction}

The initial analysis is essentially the same for each observation, and
is detailed in the following steps:

\begin{itemize}

\item For each observation, the relevant Pipeline Processing System
(PPS) products (event lists, source lists, attitude files,
housekeeping files \etc), from the standard analysis carried out at
the SSC, are collected together.

\item For each instrument, region files are created from the PPS   
source lists. These are then used to remove all the source events from
each of the relevant event files. A conservative extraction radius of
36\arcs\ is used to remove the sources (for comparison, Lumb \etal\
(2002) used 25\arcs). These regions are also removed from previously
created mask files (these are required to calculate losses in area due
to source removal).

\item A visual inspection is made of the data to make sure that there
are no strange features in the field, and to ascertain whether there
are any very bright point sources or large diffuse sources which could
contaminate the background, even after source subtraction. Datasets
which fail this inspection are rejected from any subsequent analysis.

\item Each of the event files are then filtered for periods of high
background (solar proton flares). Lightcurves are created using the
whole detector, for single events in the energy range 10$-$15\,keV
(where no source counts are expected, due to the very low effective
area of the telecope) and with FLAG values as defined by
\verb£#XMMEA_EM£ or \verb£#XMMEA_EP£. E.g.\, for PN, using the
following expression for the XMM-SAS task evselect;

\verb£PI in [10000:15000] &&#XMMEA_EP &&PATTERN==0£

Good Time Interval (GTI) files are created from these lightcurves by
applying an upper count rate threshold of 100 (PN) or 35 (M1/M2)
ct/100\,s. The event files are then filtered, keeping the low count
rate time periods. The amount of time lost due to periods of high
background flaring can be seen in fig.\,1, where the distribution of
the fraction of data left after flare-screening is shown for all the
datasets analysed.

\begin{figure}
\vspace{8.7cm}
\includegraphics{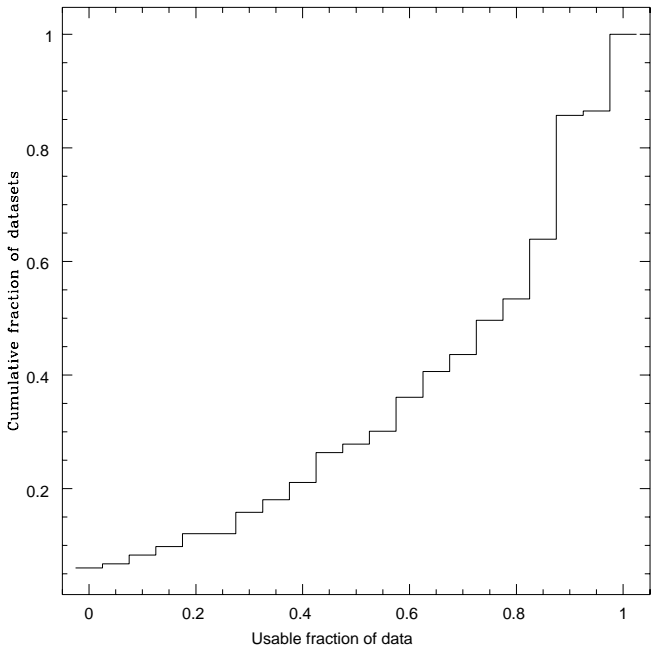}
\caption{The effects of high background flaring: The usable fraction
of data available after flare-screening is shown for all the datasets
analysed.}
\label{fig0}
\end{figure}

\item The event files are then filtered further. Events with energies
below 150\,eV are discarded. For PN, only singles and doubles are
retained, for M1/M2, singles, doubles, triples and quadruples are
retained. Finally, the event lists are filtered using the
\verb£#XMMEA_EM/P£ FLAG expressions, excepting that events from
outside the field of view (out-of-FOV) are kept.

\item In the case of PN, it was necessary to further filter the files
to eliminate a small number of persistent bad (bright) pixels/columns,
occurring in many (though not all) of the observations. The same
pixels were also found to contaminate the PN closed datasets. These
pixels were removed from all the PN datasets. Given the large-scale
(1\arcm) of the final BG maps, this had a very small effect, as the
loss in area is very small. Nevertheless, this difference was
corrected for in the final maps. The pixels removed were as follows:
CCD1 col.13 \& (56,75), CCD2 (46$-$47,69$-$72), CCD5 col.11, CCD7
col.34, CCD10 col.61, CCD11 (47$-$48,153$-$156) (50,161$-$164). No bad
pixels were removed from any of the MOS datasets (pointed observations
or closed datasets). The event files are now filtered and have had all
sources removed.

\item For each of the three instruments, a small-scale (4\arcs)
non-vignetted exposure map (with dimensions given in Table 2) is
created. From the source-removed mask file, an area map (4\arcs) is
created, containing zero values at the positions where sources have
been removed, and unity values elsewhere. These two maps are
multiplied and rebinned to form a large-scale (1\arcm; see Table 2)
`area-times-exposure' map (see Fig.2).

\begin{figure*}
\vspace{6cm}
\includegraphics{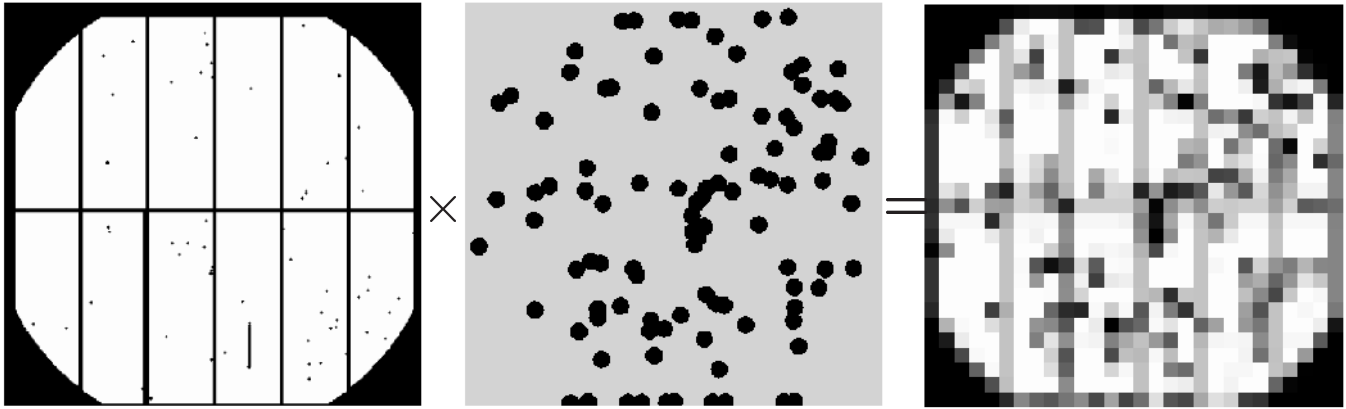}
\caption{A non-vignetted (4\arcs) exposure map is multiplied by a
source-removed (4\arcs) mask map and rebinned to create a large-scale
(1\arcm) `area-times-exposure' map (example is for PN).}
\label{fig1}
\end{figure*}

\item For each instrument (of 3) and each energy range (of 6),
(1\arcm) detector maps are created. Because X-ray photons cannot reach
the out-of-FOV areas of the detector, the events in these areas of the
detector are solely due to the instrumental and particle components of
the background. By making use of the Closed datasets, and comparing
the number of counts in the out-of-FOV regions of the Closed datasets
and the particular observation dataset in question, one can separate
(for each mode-dependent instrument and energy range) the 1\arcm\
detector map into a particle (pa) map, containing the particle and
instrumental background components and a photon (ph) map, containing
the photon component of the background. This is done, producing 2 maps
for each instrument and each energy range. 

\item The particle images are then exposure-corrected via a direct
division by the LIVETIME (corrected for periods of high background and
deadtime, deadtime being times when detector areas are affected by
cosmic rays, and therefore not available to detect X-rays), and a
division by a 1\arcm\ detector mask of the sensitive area of the CCDs
(to correct for chip gaps). The photon images are exposure-corrected
via a division by the appropriate `area-times-exposure' map.

\end{itemize}

\subsection{Data Analysis: Observation, Mode and Filter Selection}

The above preparatory analysis has been performed for 116 XMM-Newton
observations, spanning a range in instrumental modes, filters,
exposure times, and degree and duration of high-background flaring.

In order to produce the final background maps, the observations have
been grouped together in terms of instrument/mode/filter etc. We then
discarded datasets suffering from any of the following deficiencies:

\begin{itemize}

\item All observations containing a significantly bright source whose
wings could still contaminate the background after source subtraction. 

\item All observations containing a large diffuse source, which could
contribute to the estimated background. 

\item All observations where the background flaring was of such an
extent that, after flare-removal, less than 10\% of the original
exposure remained. 

\end{itemize}

The remaining 72 `clean' observations used in the production of the
final background files are summarized in Table~3, which lists the
revolution number, the RA and Dec, the Galactic hydrogen column, and a
code giving the instrumental mode and filter for, respectively, M1, M2
and PN [f - full-frame mode, e - full-frame-extended mode , t - thin
filter, m - medium filter]. Also given are a nominal M1 exposure time,
and the fraction of the exposure time remaining after removal of
high-background periods and accounting for deadtime effects (here the
PN fraction is given).

It is instructive to study some properties of the 72 clean
observations.  Fig.\,3 shows the distributions of (from left to right)
the revolution number of the observation, the Galactic hydrogen column
density, and the live exposure time, after cleaning for times of high
background.

Almost all of the observations were taken in a 100-revolution period
between the end of June 2001 and the middle of January 2002. The
Galactic hydrogen column density distribution shows that the majority
of the observations are pointed in directions of low to medium
column. A very small number of observations are pointed in directions
of very high column. The post-flare-removal exposure-weighted mean
value of the hydrogen column over all the 72 observations is
6.83$\times10^{20}$\,cm$^{-2}$. The post-flare-removal exposure times
show a systematically larger exposure time for the MOS instruments
than for the pn (the M2 distribution is essentially identical to the
M1 distribution shown). This is partly due to the pn being slightly
more sensitive to proton flaring, but is mainly due to the fact that
operational overheads generally lead to the pn having shorter exposure
times (by $\approx$5\,ks) than the MOS. Mean values of the live
post-flare-removal exposure times are: M1: 22.0\,ks (standard
deviation 13.8\,ks), M2: 22.2\,ks (s.d. 13.8\,ks), PN: 14.8\,ks
(s.d. 11.9\,ks).

\begin{figure*}
\vspace{7cm}
\includegraphics{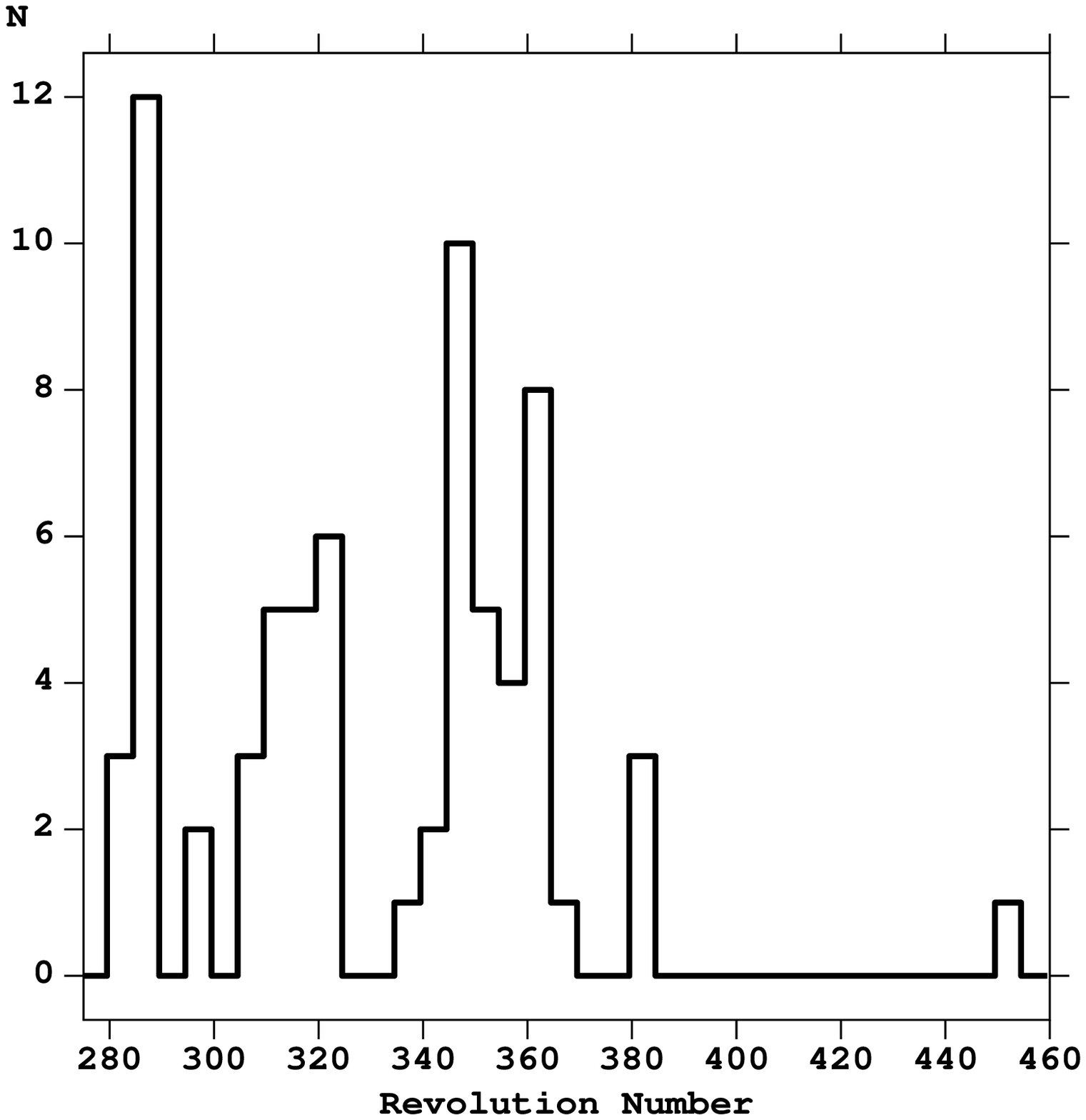}
\includegraphics{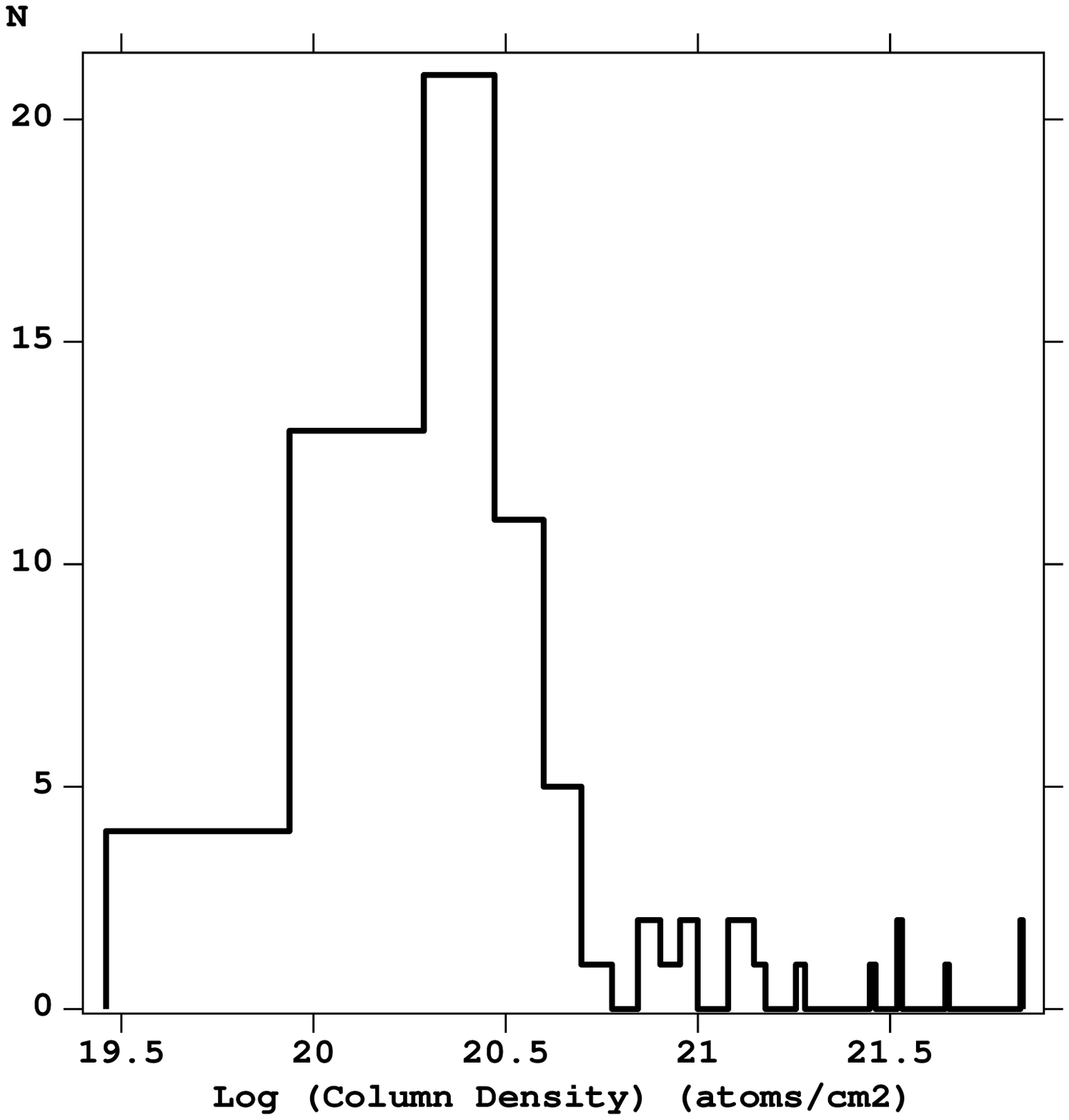}
\includegraphics{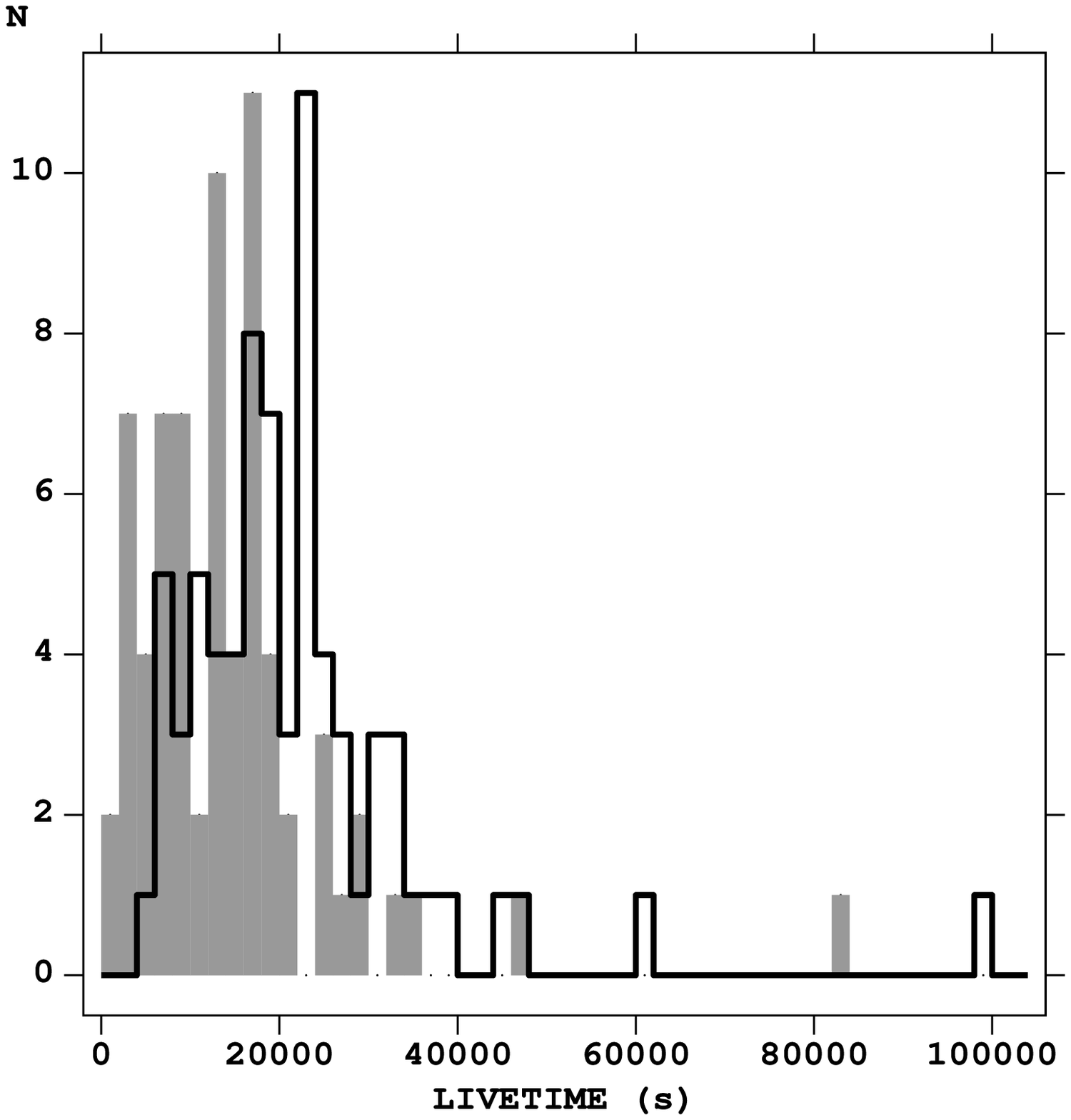}
\caption{Histogram distributions for the 72 clean observations of
(from left to right) the observation revolution number, the Galactic
hydrogen column, and the live exposure time (after cleaning for high
background times) for M1 (thick line) and PN (filled region).}
\label{fig4} 
\end{figure*} 

\begin{table*}
\vspace{15mm}
\caption[]{Summary of the final cleaned and filtered observations used in
the production of the EPIC background files. Given is the revolution
number, the J2000.0 equinox RA and Dec, and the Galactic hydrogen
column (in units of atoms cm$^{-2}$) towards these coordinates. Also
given is a code giving the instrumental mode and filter for M1, M2 and
PN [f - full-frame mode, e - full-frame-extended mode , t - thin
filter, m - medium filter], plus the nominal exposure time, and the
fraction of the pn exposure time remaining after removal of
high-background periods. }
\begin{tabular}{|cccccrc|cccccrc|}
\hline
Rev & RA & Dec & nH & Mode/Filt. & Exp. & f(Ex) & 
Rev & RA & Dec & nH & Mode/Filt. & Exp. & f(Ex) \\ 
    & \multicolumn{2}{c}{(2000.0)} & $10^{20}$& M1M2PN & (s) & 
&   & \multicolumn{2}{c}{(2000.0)} & $10^{20}$& M1M2PN & (s) & \\ \hline

281 &164.26 & -3.65 & 3.57 & ft\,\,ft\,\,et\,\, & 40260 & 0.61 & 324 & 81.26 & -33.69 & 2.22& ft\,\,ft\,\,ft\,\, & 27786 & 0.65 \\ 
283 &16.93 & -17.49 & 1.53 & fmfmfm & 11813 & 0.13             & 339 &125.14 & 21.10 & 4.23 & fmfmfm & 21617 & 0.88 \\ 
284 &20.15 & -10.94 & 3.28 & ft\,\,ft\,\,ft\,\, &  9900 & 0.40 & 343 &103.37 & -23.69 & 18.7 & fmfmem & 47559 & 0.37 \\ 
285 &157.69 & -46.36 & 14.9 & fmfmem &  9215 & 0.89             & 344 &121.11 & 65.02 & 4.32 & fmfmfm & 22615 & 0.73 \\ 
286 &187.81 & 20.77 & 2.01 & ft\,\,fmet\, & 24211 & 0.81       & 345 &144.97 & 35.93 & 1.43 & fmfmem & 34326 & 0.45 \\ 
286 &187.81 & 21.27 & 1.84 & ft\,\,fmet\, & 24211 & 0.89       & 345 &121.38 & 24.86 & 3.84 & ft\,\,ft\,\,et\, & 24326 & 0.70 \\ 
286 &187.81 & 21.77 & 1.76 & ft\,\,fmet\, & 24214 & 0.66       & 346 &94.45 & -32.89 & 4.19 & ft\,\,ft\,\,et\, & 11773 & 0.90 \\
287 &35.81 & -4.82 & 2.65 & ft\,\,ft\,\,et\, & 24218 & 0.89   & 346 &93.85 & -33.39 & 3.43 & ft\,\,ft\,\,et\, & 14772 & 0.46 \\ 
287 &36.31 & -5.15 & 2.67 & ft\,\,ft\,\,et\, & 24318 & 0.89   & 346 &93.85 & -32.89 & 3.48 & ft\,\,ft\,\,et\, & 14771 & 0.71 \\ 
287 &35.98 & -5.15 & 2.64 & ft\,\,ft\,\,et\, & 28512 & 0.88   & 346 &127.91 & 52.78 & 3.85 & fmfmfm & 16916 & 0.90 \\ 
287 &35.64 & -5.15 & 2.58 & ft\,\,ft\,\,et\, & 24313 & 0.87   & 347 &324.52 & -43.72 & 2.91 & ft\,\,ft\,\,ft\,\, &  7615 & 0.91 \\ 
288 &352.95 & 19.65 & 4.19 & fmfmfm & 11513 & 0.77             & 348 &154.66 & 41.45 & 1.12 & ft\,\,ft\,\,ft\,\, & 29617 & 0.76 \\ 
288 &36.98 & -5.15 & 2.64 & ft\,\,ft\,\,et\, & 25149 & 0.74   & 348 &137.86 & 5.88 & 3.65 & ft\,\,ft\,\,et\, & 20072 & 0.22 \\ 
288 &36.64 & -5.15 & 2.66 & ft\,\,ft\,\,et\, & 12268 & 0.17   & 349 &163.18 & 57.51 & 0.56 & fmfmfm & 37950 & 0.77 \\ 
288 &14.69 & -27.58 & 1.91 & ft\,\,ft\,\,et\, &  9209 & 0.62   & 353 &162.47 & 33.01 & 1.98 & ft\,\,ft\,\,et\, & 37800 & 0.36 \\ 
299 &53.08 & -27.79 & 0.90 & ft\,\,ft\,\,et\, & 39765 & 0.49   & 353 &152.79 & 55.78 & 0.78 & xx\,\,ft\,\,ft\,\, & 32615 & 0.83 \\ 
299 &53.09 & -27.79 & 0.90 & ft\,\,ft\,\,et\, & 56698 & 0.65   & 353 &143.95 & 61.38 & 2.68 & fmfmet\, & 34371 & 0.89 \\ 
300 &213.95 & 11.47 & 1.82 & ft\,\,ft\,\,et\, & 25719 & 0.88   & 354 &335.21 & -24.70 & 1.94 & fmfmfm & 27617 & 0.85 \\ 
305 &222.38 & 8.97 & 2.03 & ft\,\,ft\,\,ft\,\, & 40387 & 0.86 & 354 &314.11 & -4.65 & 4.97 & xx\,\,fmem & 16770 & 0.89 \\ 
307 &201.70 & -47.50 & 9.54 & fmfmfm & 39978 & 0.47             & 355 &333.91 & -17.75 & 2.36 & ft\,\,ft\,\,xx\,\, & 34397 & 0.96 \\ 
308 &36.65 & -5.15 & 2.63 & ft\,\,ft\,\,et\, & 12666 & 0.89   & 355 &333.91 & -17.75 & 2.36 & ft\,\,ft\,\,ft\,\, & 10824 & 0.62 \\
310 &64.90 & 56.02 & 44.6 & fmfmfm & 32310 & 0.40             & 356 &333.90 & -17.75 & 2.36 & ft\,\,ft\,\,ft\,\, &109427 & 0.77 \\ 
311 &224.60 & -31.70 & 8.42 & ft\,\,ft\,\,ft\,\, & 52799 & 0.43 & 359 &148.42 & 1.61 & 3.53 & ft\,\,ft\,\,et\, & 10020 & 0.39 \\ 
312 &0.67 & 62.77 & 69.3 & fmfmsm & 33000 & 0.49             & 360 &152.77 & -4.66 & 3.83 & fmfmem & 19130 & 0.21 \\ 
312 &4.07 & 81.58 & 13.9 & fmfmfm & 39337 & 0.35             & 360 &169.57 & 7.79 & 3.51 & fmfmft\,\, & 62616 & 0.78 \\ 
314 &94.65 & 78.38 & 7.30 & fmfmft\,\, & 26847 & 0.40         & 360 &182.38 & 43.71 & 1.35 & ft\,\,ft\,\,em & 14317 & 0.87 \\ 
316 &259.92 & -25.04 & 28.5 & ft\,\,ft\,\,ft\,\, & 12225 & 0.70 & 361 &355.00 & -12.31 & 2.76 & ft\,\,ft\,\,et\, & 14369 & 0.86 \\ 
316 &253.52 & -39.88 & 69.1 & fmfmfm & 39820 & 0.46             & 361 &339.27 & -15.29 & 3.89 & ft\,\,ft\,\,et\, & 24370 & 0.78 \\ 
317 &58.46 & 23.41 & 9.47 & ft\,\,ft\,\,ft\,\, & 34335 & 0.31 & 362 &346.21 & 3.18 & 5.25 & ft\,\,ft\,\,ft\,\, & 12566 & 0.88 \\ 
317 &51.39 & 30.75 & 13.8 & ft\,\,ft\,\,ft\,\, & 32725 & 0.32 & 363 &159.97 & 20.87 & 2.02 & ft\,\,ft\,\,et\, & 14365 & 0.47 \\ 
317 &76.31 & -28.80 & 1.49 & ft\,\,ft\,\,et\, & 48870 & 0.64   & 364 &183.06 & 13.23 & 2.57 & ft\,\,ft\,\,em & 22867 & 0.86 \\ 
321 &250.21 & -24.13 & 12.5 & ft\,\,ft\,\,et\, & 20121 & 0.88   & 366 &231.50 & 51.64 & 1.56 & ft\,\,ft\,\,ft\,\, & 36926 & 0.72 \\ 
321 &250.21 & -24.43 & 12.4 & ft\,\,fmet\, & 18626 & 0.82       & 383 &36.02 & -3.85 & 2.49 & ft\,\,ft\,\,et\, & 13382 & 0.85 \\ 
322 &79.07 & 79.69 & 7.99 & fmfmfm & 31997 & 0.90             & 383 &35.52 & -3.52 & 2.48 & ft\,\,ft\,\,et\, & 13382 & 0.44 \\ 
323 &263.57 & -26.11 & 33.3 & ft\,\,ft\,\,ft\,\, & 24618 & 0.61 & 383 &35.86 & -3.52 & 2.52 & ft\,\,ft\,\,et\, & 12380 & 0.87 \\ 
323 &263.57 & -26.11 & 33.3 & ft\,\,ft\,\,ft\,\, & 25115 & 0.60 & 451 &187.65 & 41.62 & 1.78 & fmfmem & 17521 & 0.85 \\ \hline
\end{tabular}
\end{table*}

Table~4 summarizes the final cleaned observation information in terms
of the different combinations of instrument, instrument mode and
filter used.  The nominal exposure time is the sum of the individual
LIVETIMEs \ie\ corrected for periods of high-background and deadtime
(see however the discussion of the exposure maps in Section\,5.1).

\begin{table*}
\caption[]{Summary of the cleaned and filtered observations used in
the production of the EPIC background files, separated into the
different combinations of instrument, instrument mode and filter
used. The exposure time is the sum of the individual LIVETIMEs \ie\
corrected for periods of high-background and deadtime. The
post-flare-removal exposure-weighted mean value of the hydrogen column
is also tabulated. }
\begin{tabular}{lllrrr}
\hline
Instrument & Mode                & Filter & Number of    & Exposure   & nH \\
           &                     &        & Observations &   Time (s) & $10^{20}$\,cm$^{-2}$ \\ \hline
 MOS1      & Full-Frame          & Thin   & 49           &    1055905 & 4.49 \\
 MOS1      & Full-Frame          & Medium & 21           &     488422 & 12.7 \\
 MOS2      & Full-Frame          & Thin   & 46           &    1004709 & 4.40 \\
 MOS2      & Full-Frame          & Medium & 26           &     592975 & 11.1 \\
 PN        & Full-Frame          & Thin   & 18           &     351549 & 6.35 \\
 PN        & Full-Frame          & Medium & 12           &     188159 & 13.3 \\
 PN        & Full-Frame-Extended & Thin   & 32           &     416739 & 3.05 \\
 PN        & Full-Frame-Extended & Medium &  8           &      82957 & 6.13 \\ \hline
\end{tabular}
\end{table*}

\subsection{Data Analysis: Cubes and Clipping}

Here we describe the production of a single `averaged' background map
from many background maps over different observations. At any given
position in DETX/Y coordinates, a subset of the datasets used may be
affected by contamination or diffuse sources. Such effects can be
removed by clipping outliers, before taking the mean of the remaining
`good' images at this position.

In order to produce a particular set of background maps (in the
different energy bands) for a particular instrument/mode/filter
combination, the following steps are taken. Let us take for example
the first entry in table~4, \ie\ MOS1, full-frame mode, thin filter.

\begin{itemize}

\item For each of the 49 relevant observations, comprising in total
over 1\,Msec of clean, low-background data, the 0-band (full energy
band) 1\arcm\ photon detector maps of the background are stacked
together into a 3-D `imagecube' (see Fig.\,4) of dimensions DETX, DETY
(each in the case of MOS1 of size 34; see Table~2) and N$_{\mbox
{\small obs}}$ (of size, in this example, 49).

\begin{figure*}
\vspace{9cm}
\includegraphics{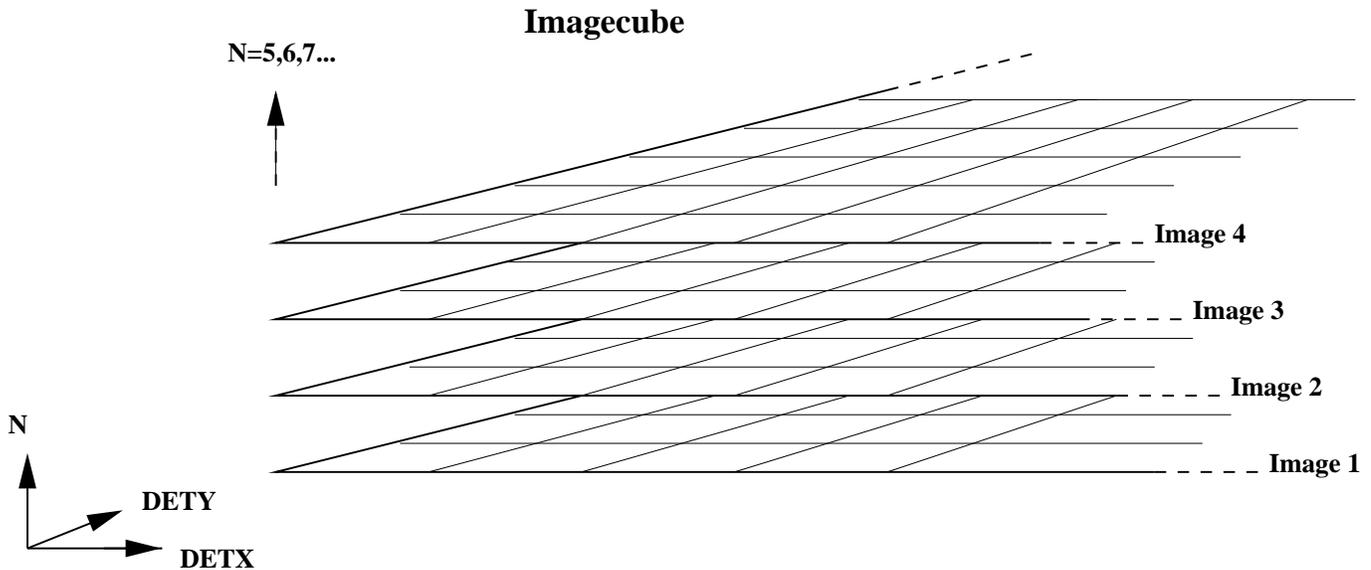}
\caption{A 3-D `imagecube' of dimensions DETX, DETY and N$_{\mbox
{\small obs}}$.}
\label{fig2}
\end{figure*}

\item For each DETX/DETY, a statistical analysis of the (in this case,
49) values at that particular DETX/DETY point is performed. Here, a
`clipcube' is constructed (with dimensions identical to the input
imagecube), containing information (1's and 0's) as to which cells in
the imagecube are within the allowed range and which values are not,
\ie\ which values are `clipped' (see Fig.~5). The allowable range is
defined as within some number of standard deviations ($\sigma$s) from
the mean value at that DETX/DETY. For the present analysis, this
number of $\sigma$s is set at 1.2. Steps are taken, including the
prior removal of very strongly aberrant pixels, to ensure that the
initial mean used to define the clip limits is not seriously biased by
outliers. The mean value at that particular DETX/DETY is then
calculated from the remaining `$\sigma$-clipped' values.

\begin{figure*}
\vspace{9cm}
\includegraphics{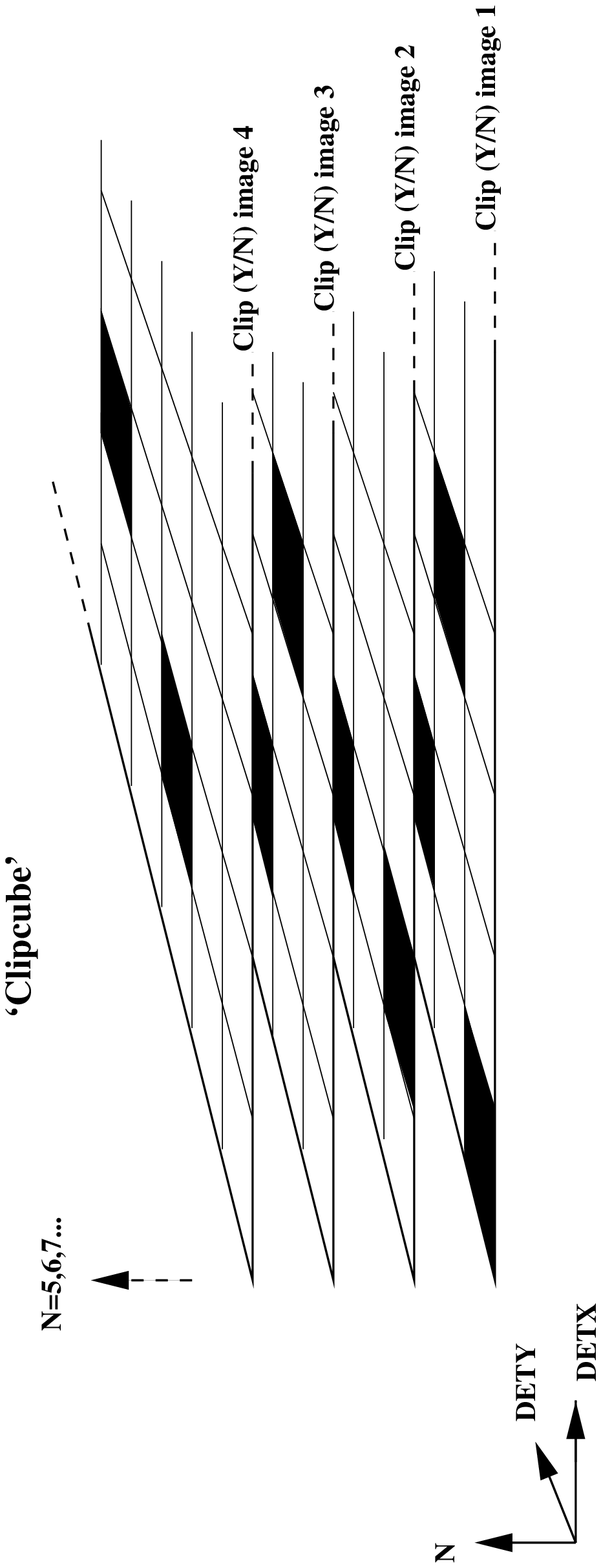}
\caption{A 3-D `clipcube' of dimensions DETX, DETY and N$_{\mbox
{\small obs}}$. In this example, for DETX/DETY=(1,1), the value in
image 1 is outside of the allowable range, and is to be clipped from
the calculation of the mean value for DETX/DETY=(1,1).}
\label{fig3}
\end{figure*}

\item Imagecubes are then created, as in the 0-band, from the
intermediate energy band (bands 1$-$5) 1\arcm\ photon detector maps of
the background. Due to the limited statistics in the band 1$-$5
imagecubes, the broad-band clipcube is used to create the
$\sigma$-clipped band 1$-$5 images. The same procedure is performed
for the particle images, and for all instrument/mode/filter
combinations.

\end{itemize}

\section{The Background Maps}

$\sigma$-clipped, and averaged, exposure-corrected background maps
have been created as described above for each instrument/mode/filter
combination analysed (of 8) and in each energy band (of 6). Photon and
particle maps have been created separately, as have maps with the two
components recombined. This leads to a total of 144 background maps,
and these are shown in Figs\,6$-$9. The maps can be obtained from
\verb£http://www.sr.bham.ac.uk/xmm3/£ 

\verb£A1_ft0000_cphim4M1.fits£ is, as an example of the file naming,
an exposure-corrected photon background map. A particle map is named
\verb£*cpaim*£, instead of \verb£*cphim*£, and a particle and photon
combined map \verb£*cim*£. The example above is a MOS1 map (\verb£M1£
instead of \verb£M2£ or \verb£PN£), and is in energy band \verb£4£ (of
\verb£0£$-$\verb£6£). The mode is full-frame (given by the \verb£f£),
as opposed to (for PN) extended full-frame mode (\verb£e£), and the
filter is thin (\verb£t£), as opposed to medium (\verb£m£). The six
character mode+filter string is as in Table~3, hence the corresponding
PN file is named \verb£A1_0000ft_cphim4PN.fits£. The \verb£A1£
indicates a first release of these background maps.

\begin{figure*}
\vspace{16.5cm}
\includegraphics{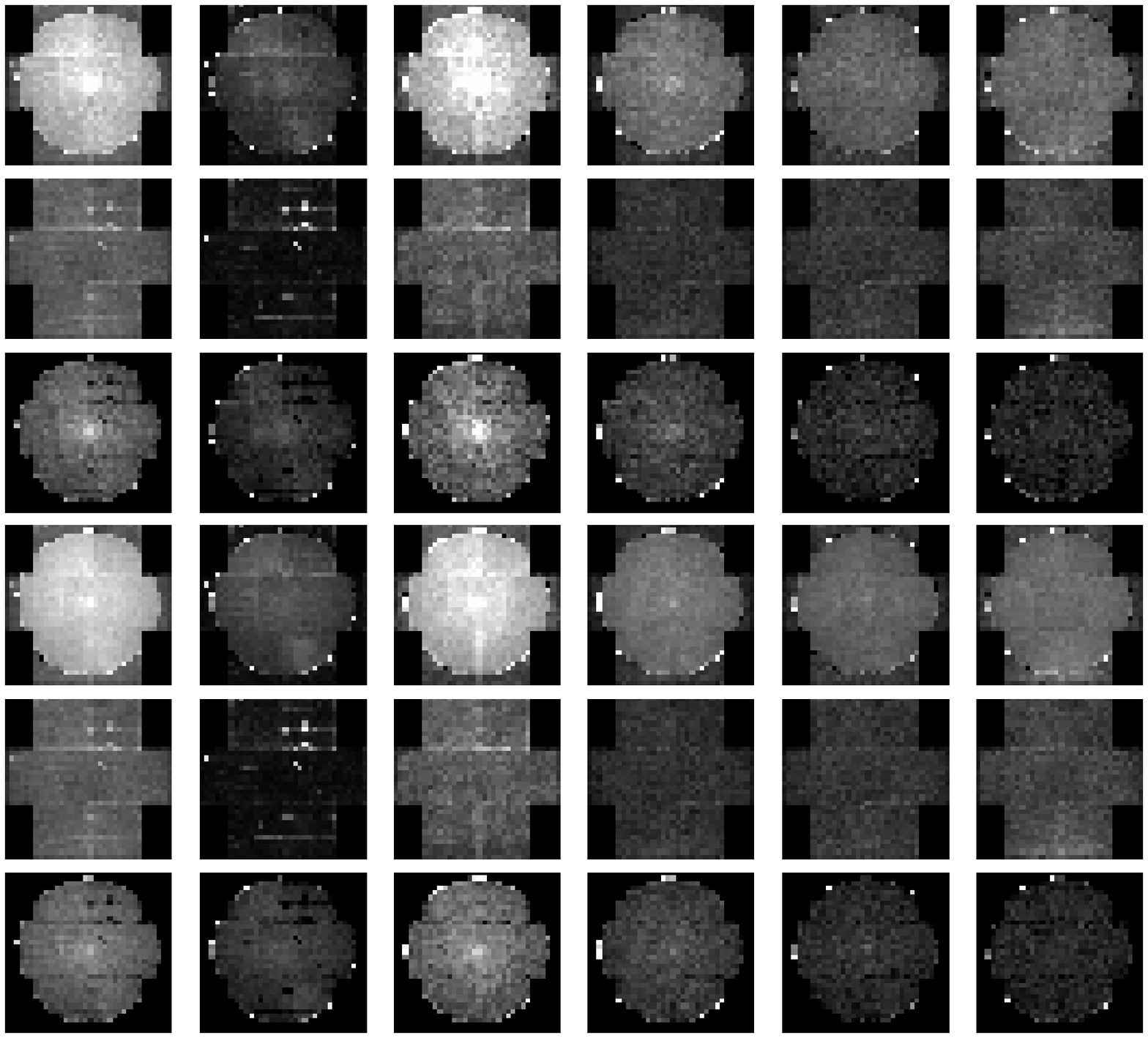}
\caption{The final $\sigma$-clipped, averaged, and exposure-corrected
background maps, for MOS1, full-frame mode. Energy band 0 (full-band),
then bands 1$-$5 run from left to right. Rows 1$-$3 are for medium
filter and show (row\,1) the particle and photon combined (cim) maps,
(row\,2) the particle (cpaim) maps, and (row\,3) the photon (cphim)
maps. Rows 4$-$6 show the equivalent maps for thin filter. All
0$-$band maps are to the same level of scaling. All other maps are to
a scaling 1/3 that of the 0$-$band maps. Note that at the very edges
of the photon (and combined) map FOVs, there are sometimes a few
pixels with unusually large or small values. This is due to extremely
small exposure values in the original observation maps amplifying the
noise.}
\label{fig}
\end{figure*}

\begin{figure*}
\vspace{16.5cm}
\includegraphics{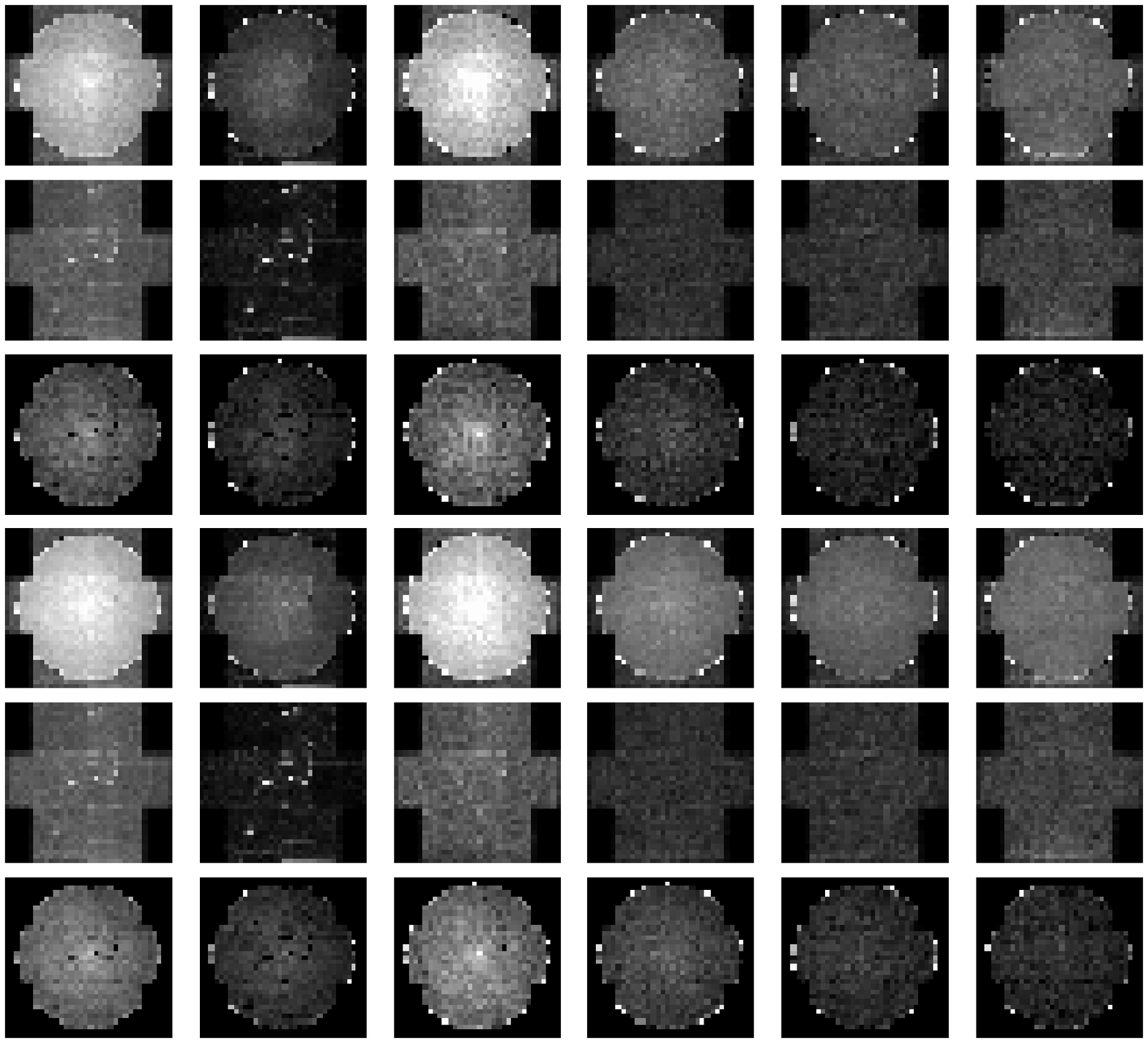}
\caption{The final $\sigma$-clipped, averaged, and exposure-corrected 
  background maps, as for Fig.\,6, but for MOS2, full-frame mode.}
\label{fig}
\end{figure*}

\begin{figure*}
\vspace{16.5cm}
\includegraphics{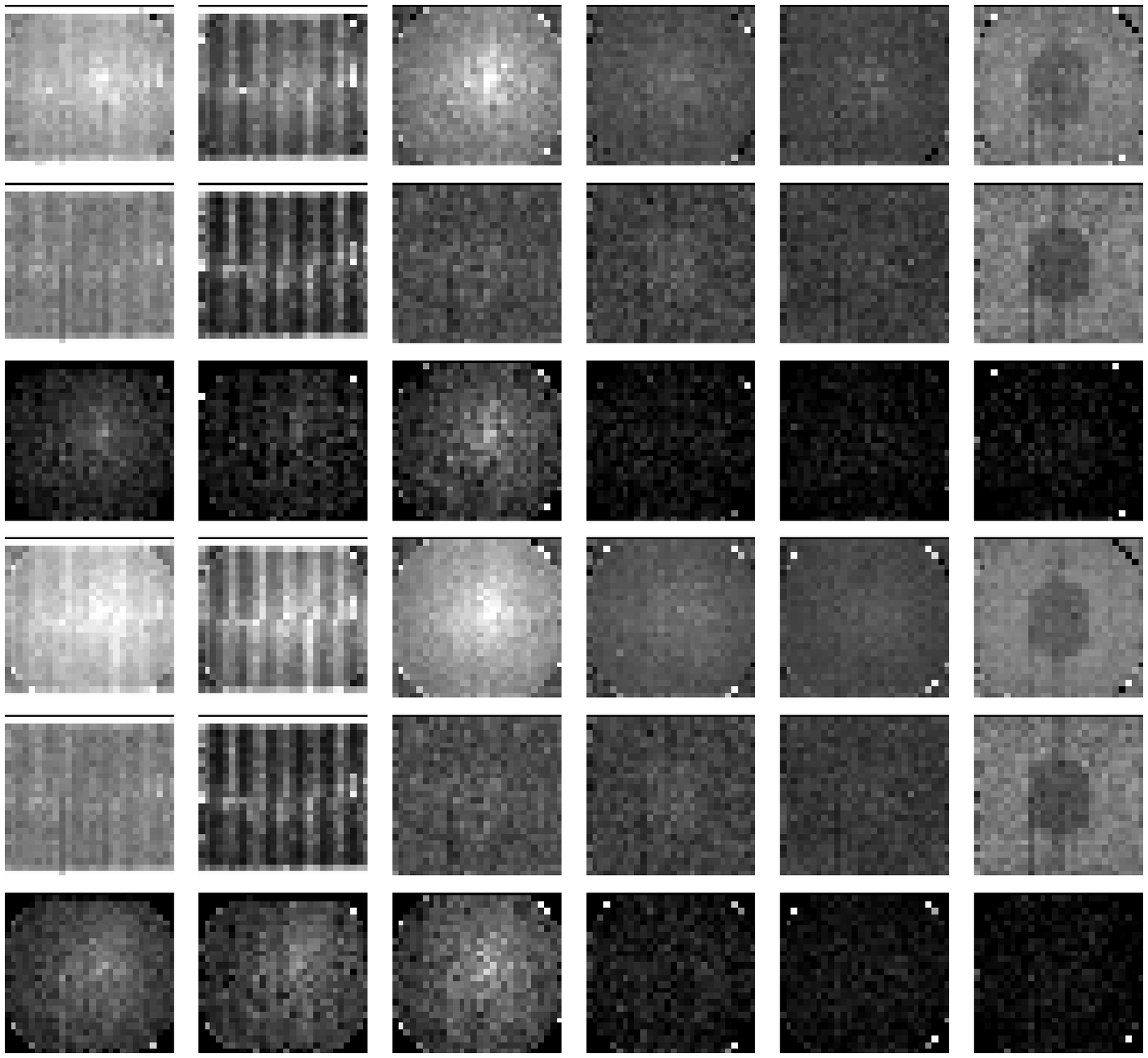}
\caption{The final $\sigma$-clipped, averaged, and exposure-corrected 
  background maps, as for Fig.\,6, but for pn, full-frame mode.}
\label{fig}
\end{figure*}

\begin{figure*}
\vspace{16.5cm}
\includegraphics{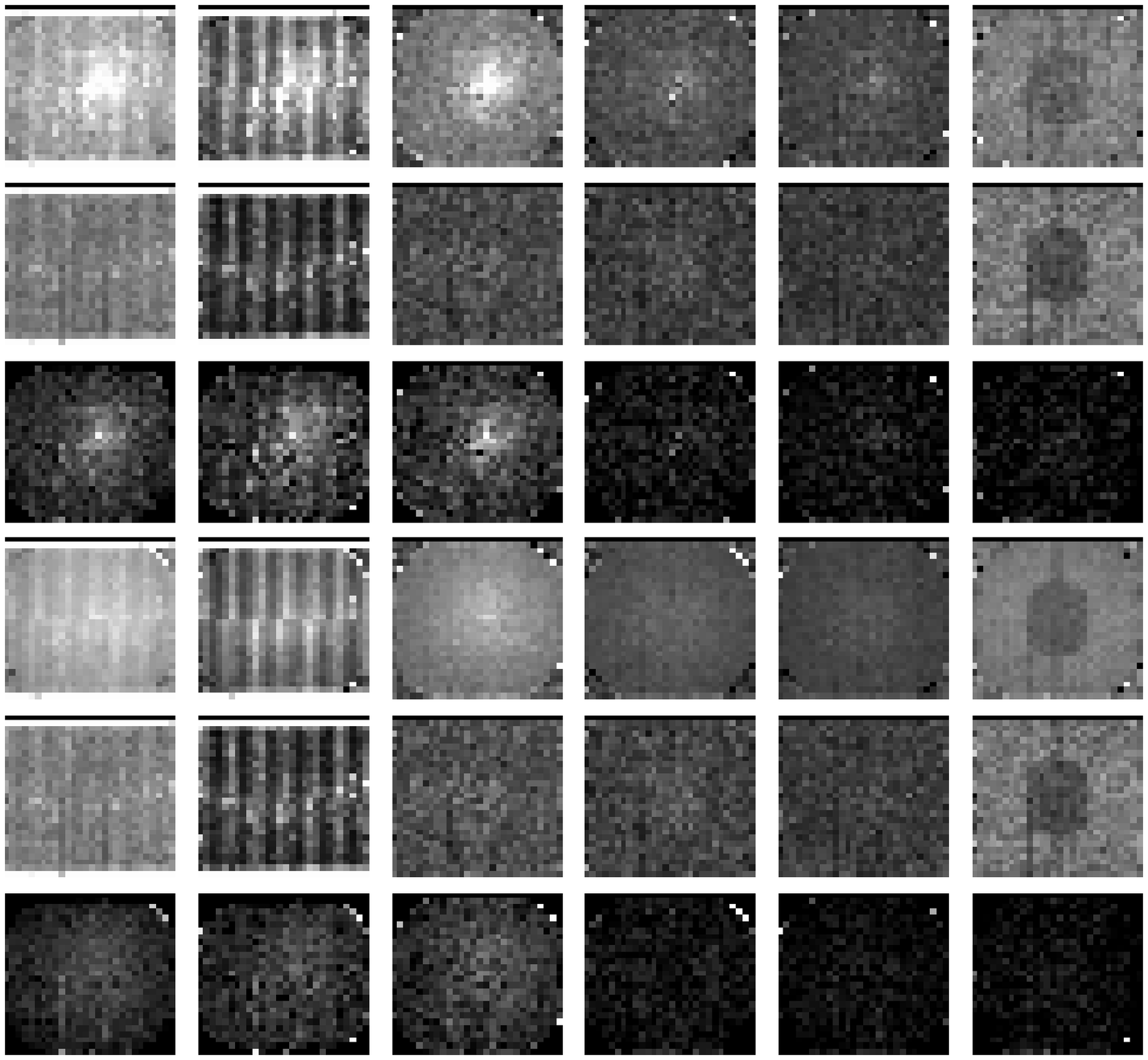}
\caption{The final $\sigma$-clipped, averaged, and exposure-corrected 
  background maps, as for Fig.\,6, but for pn, extended full-frame mode.}
\label{fig}
\end{figure*}

\subsection{The Background Maps: Usage}

When might an observer with their own EPIC data want to make use of
these background maps? In order to analyse extended objects, and make
background-subtracted images, estimate low-surface brightness flux
levels, create radial profiles or perform 1- or 2-D surface brightness
fitting, one needs a map of the appropriate background. Sometimes
however, the diffuse, extended nature of the user's target source is
such that the determination of the background from their own dataset
is difficult or impossible. Also, using a background from a
significantly spatially removed section of the same data leads to
problems with vignetting and detector variations. Hence the need for
the independently-produced background maps created here.
 
What follows is a recipe for how an observer can make use of the background
maps in conjunction with their own source data. Caveats, and problems that
can occur when the recipe is not or can not be followed are given
thereafter.

\begin{itemize}
  
\item The source data should be flare-rejected to a similar level to
that applied in the creation of the background maps. $^1$
  
\item Images should be created from the flare-rejected source datasets
in the same XID/PPS energy bands (bands 0$-$5) as used here. $^2$
  
\item As a particular source image will be created (most commonly) in
sky coordinates, and will have a resolution finer than 1\arcm, the
routine {\em BGrebinimage2SKY} should be used to rebin and sky-project
the equivalent background map(s) to the resolution and sky position of
the source image (see Sect.\,4.2).
  
\item Depending on whether the source maps are raw count maps or
exposure-corrected flux maps, the equivalent rebinned background
map(s) (which are exposure-corrected) may be scaled appropriately,
either by an exposure time, or using an appropriate exposure map.
  
\item A comparison of the counts or flux in the out-of-FOV areas of
the source image and background map yields the scaling by which the
particle component (the `cpaim' map) of the background needs be
scaled. $^3$
  
\item If source-free regions within the FOV exist within the source
dataset, then count/flux comparisons in these regions yield the
scaling for the photon component (the `cphim' map). The scaled photon
component can then be add to the scaled particle component to give the
final background map. $^3$

\end{itemize}

$^1$ If such a flare-rejection method as used here leads to zero or
very few Good Times, then the user's data is heavily
flare-contaminated, and the background maps are therefore not suitable
for source data extracted from the whole dataset. Small discrepancies
due to slightly different flare-rejection methods may be accounted for
by the subsequent scaling.

$^2$ Though the user should work using the same energy bands (bands
0$-$5) as used in the present analysis, note that the standard XID/PPS
bands have been used, and that extension to larger energy bands can
easily be performed by summing individual band images. For example,
band 1 is heavily contaminated by detector noise in the particle
background, so the user may prefer to work in band (2+3+4+5).

$^3$ If source-free regions within the FOV do exist within the source
dataset, then the user is advised to work as above with the particle
and photon (`cpaim' \& `cphim') maps separately. The particle and
photon scaling factors may be different (hence the fact that separate
particle and photon background maps have been made available). If no
source-free regions exist, the user may be forced to assume that the
scalings are the same. Here, just the `cim' images need be used. A
script for comparing out-of-FOV counts (from event files), {\em
compareoutofFOV}, is available on http://www.sr.bham.ac.uk/xmm3/. Note
especially that at the very edges of the map FOVs, there are a few
pixels with unusually large and small values. This is due to extremely
small exposure values in the original observation maps amplifying the
noise. These areas should be avoided when calculating scaling values.

\subsection{The Background maps: Statistics}

For each of the instrument/mode/filter combinations, and for each of
the energy bands 0$-$5, mean count rates (in units of ct ks$^{-1}$
arcmin$^{-2}$) and standard deviations about these mean values have
been calculated using all of the contributing observations listed in
Table 3. This has been done separately for the photons and for
particles. For the photons, the mean values were calculated within the
central 16\arcm $\times$16\arcm, to avoid the small number of pixels
discussed above with unusually large and small values. Table 5 lists
these values for the photons, table 6 lists them for the particles.

\begin{table*}
\caption[]{Mean count rates for the {\em photon} background maps (over
the central 16\arcm $\times$16\arcm ), for the different instrument,
mode and filter combinations, and for each of the six (five plus
total) standard energy bands.}
\begin{tabular}{llrcccccc}
\hline
Instr. & Mode/  & N$_{\rm obs}$ & \multicolumn{6}{c}{Mean Count Rate (ct ks$^{-1}$ arcmin$^{-2}$) (+ standard 
  deviation)} \\
       & filter & & Band {\bf 0:} & Band {\bf 1:} & Band {\bf 2:} & Band {\bf 3:} & Band {\bf 4:} & Band {\bf 5:} \\
       & & & 200$-$12000\,eV & 200$-$500\,eV & 500$-$2000\,eV & 2000$-$4500\,eV & 4500$-$7500\,eV & 7500$-$12000\,eV  
  \\ \hline 

 MOS1 & ft & 49 & 2.10 (1.36) & 0.37 (0.18) & 0.78 (0.46) & 0.49 (0.37) & 0.35 (0.22) & 0.31 (0.18) \\
 MOS1 & fm & 21 & 2.00 (1.06) & 0.32 (0.12) & 0.80 (0.38) & 0.48 (0.27) & 0.33 (0.16) & 0.29 (0.14) \\
 MOS2 & ft & 46 & 2.23 (1.40) & 0.39 (0.19) & 0.84 (0.50) & 0.53 (0.39) & 0.37 (0.22) & 0.32 (0.18) \\
 MOS2 & fm & 26 & 1.81 (1.07) & 0.31 (0.12) & 0.76 (0.39) & 0.42 (0.27) & 0.29 (0.16) & 0.27 (0.13) \\
 PN   & ft & 18 & 6.50 (3.89) & 1.90 (1.02) & 2.46 (1.28) & 0.93 (0.71) & 0.69 (0.43) & 0.61 (0.24) \\
 PN   & fm & 12 & 4.31 (2.24) & 1.13 (0.50) & 2.04 (0.94) & 0.72 (0.33) & 0.64 (0.36) & 0.68 (0.48) \\
 PN   & et & 32 & 5.19 (4.38) & 1.87 (2.06) & 1.71 (1.01) & 0.86 (0.68) & 0.72 (0.52) & 0.79 (0.55) \\
 PN   & em &  8 & 5.41 (2.66) & 1.94 (0.77) & 1.90 (0.47) & 0.90 (0.60) & 0.76 (0.50) & 0.74 (0.40) \\

\hline
\end{tabular}
\end{table*}

\begin{table*}
\caption[]{Mean count rates for the {\em particle} background maps for
the different instrument, mode and filter combinations, and for each
of the six (five plus total) standard energy bands.}
\begin{tabular}{llrcccccc}
\hline
Instr. & Mode/  & N$_{\rm obs}$ & \multicolumn{6}{c}{Mean Count Rate (ct ks$^{-1}$ arcmin$^{-2}$) (+ standard 
  deviation)} \\
       & filter & & Band {\bf 0:} & Band {\bf 1:} & Band {\bf 2:} & Band {\bf 3:} & Band {\bf 4:} & Band {\bf 5:} \\
       & & & 200$-$12000\,eV & 200$-$500\,eV & 500$-$2000\,eV & 2000$-$4500\,eV & 4500$-$7500\,eV & 7500$-$12000\,eV  
  \\ \hline 

 MOS1 & ft & 49 & 1.40 (0.11) & 0.12 (0.03) & 0.44 (0.04) & 0.24 (0.02) & 0.26 (0.02) & 0.34 (0.03) \\ 
 MOS1 & fm & 21 & 1.43 (0.11) & 0.13 (0.03) & 0.45 (0.04) & 0.24 (0.02) & 0.26 (0.02) & 0.34 (0.02) \\ 
 MOS2 & ft & 46 & 1.34 (0.09) & 0.14 (0.02) & 0.42 (0.03) & 0.23 (0.02) & 0.24 (0.02) & 0.32 (0.02) \\ 
 MOS2 & fm & 26 & 1.31 (0.09) & 0.13 (0.02) & 0.42 (0.04) & 0.23 (0.02) & 0.24 (0.02) & 0.31 (0.02) \\ 
 PN   & ft & 18 & 8.37 (2.29) & 2.13 (0.43) & 1.95 (1.36) & 1.50 (0.86) & 1.13 (0.31) & 2.05 (0.21) \\ 
 PN   & fm & 12 & 8.16 (1.60) & 2.23 (0.31) & 1.55 (0.77) & 1.32 (0.47) & 1.11 (0.20) & 2.08 (0.25) \\ 
 PN   & et & 32 & 7.96 (1.52) & 2.10 (0.29) & 1.61 (0.86) & 1.32 (0.58) & 1.15 (0.32) & 2.01 (0.20) \\ 
 PN   & em &  8 & 8.22 (2.72) & 2.27 (0.87) & 1.58 (0.81) & 1.31 (0.64) & 1.12 (0.31) & 2.05 (0.26) \\ 

\hline
\end{tabular}
\end{table*}

Several features are evident from the tables. For a particular
instrument, the photon and particle background count rates are of the
same order, the values for the pn being a factor of a few greater than
for the MOS.

Whereas, for the photons, the decline in intensity (in ct ks$^{-1}$
arcmin$^{-2}$ keV$^{-1}$ with increasing energy is quite steep, for
the particles, it is far flatter, emphasizing the points made in
Table~1, \ie\ that the particle component of the background appears
essentially flat, and dominates above 5\,keV over the photon
($\sim1.4$ power law + soft thermal lines) component. This can be seen
in Figs.\,6$-$9 also; at low energies, the photon maps are brighter,
whilst at higher energies, the particle maps dominate. In studying the
actual values, one can see that, on average, for the MOS, the photons
dominate the background over most of the spectrum, the particles only
attaining a comparable count rate in the highest energy band. For the
pn however, the particles attain a comparable count rate to the
photons at a lower energy, and are then far more dominant at the
highest energies. Again, this can be seen in Figs.\,6$-$9.

The standard deviation in the mean values represents the scatter in
the mean background count rates over the different observations. Note
though there is likely some contribution from errors introduced in the
separating of particles from photons, a large contributor being the
presence of the extremely small exposure values in the original
observation maps amplifying the noise in the out-of-FOV regions. This
effect is more prevalent in the pn datasets (where the out-of-FOV
regions are smaller). Also, there will be some contribution to the
scatter due to the Poissonian variation in the out-of-FOV counts used
to normalise the particle contribution. For the lower-exposure
datasets, this can approach 10\% in some energy bands.

As regards the particles, the scatter seen in the MOS particle
background rates is small, about 10\%, and appears roughly constant
with energy. The pn experiences a larger scatter of about 20\%, and of
this, the greatest scatter (50\%) is observed at lower energies. Not
all of this scatter however, is thought due to variations in the
actual particle count rates (see above). The photons show quite a
range in count rate scatter (from 30$-$60\% and beyond), with no
particular trend with energy or instrument.

Fig.\,10 shows an example of the distribution of background count rates
observed within the sample. Shown are (thick line) the M1 0-band
full-frame thin filter photon count rate distribution [49 entries;
Table 4] (the equivalent M2 distribution appears almost identical),
and (filled region) the PN 0-band full-frame-extended thin filter
photon count rate distribution [32 entries]. The pn distribution has a
mean value (in this example) some $\approx2.5$ times that for the MOS,
and has quite a larger scatter.

\begin{figure}
\vspace{9cm}
\includegraphics{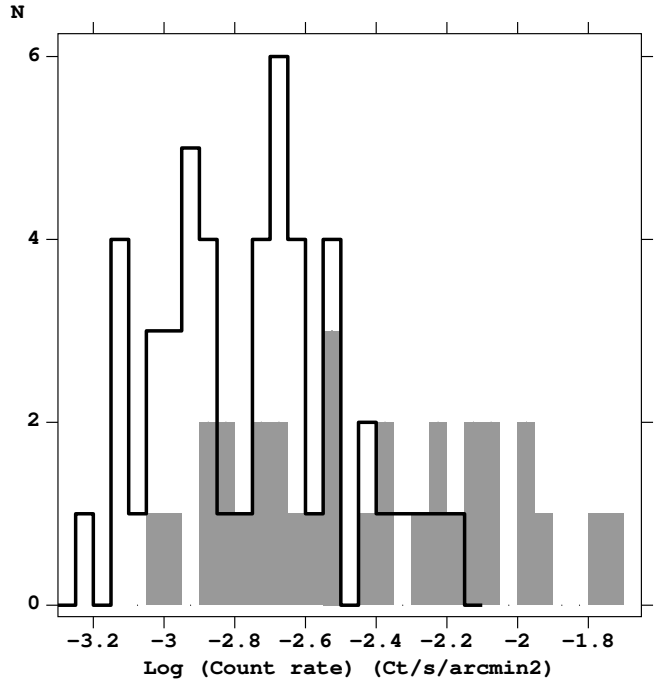}
\caption{(Thick line) the M1 0-band full-frame thin filter photon
count rate distribution, and (filled region) the PN 0-band
full-frame-extended thin filter photon count rate distribution}
\label{fig}
\end{figure}

\subsection{The Background maps: Sky coordinates, Rebinning and Reprojecting}

Usually, of course, the user will work in sky coordinates (RA \& Dec),
not detector coordinates, and to a much finer resolution than
1\arcm. With that in mind, software has been developed to rebin and
reproject onto the sky any provided background map (scaled or
otherwise) to the spatial scale and sky position of a user-input
image.

BGrebinimage2SKY is a shell script plus FORTRAN routine to convert the
low-resolution DETX/DETY background maps into high-resolution sky
(X/Y) images by interpolation. The user gives a template image
containing the attitude information and the pixel numbers and sizes,
and an associated event file (ideally the one used to create the
template image file) for general header purposes. A rebinned
background map is produced with the same resolution and at the same
sky position as the template image. The radius of the circle over
which interpolation is performed can be given. (note that the
background DETX/DETY maps are of 1\arcm\ resolution, so an
interpolation radius of this or slightly larger is recommended). An
example of the task in use can be seen in Fig.11.

\begin{figure*}
\vspace{17cm} \includegraphics{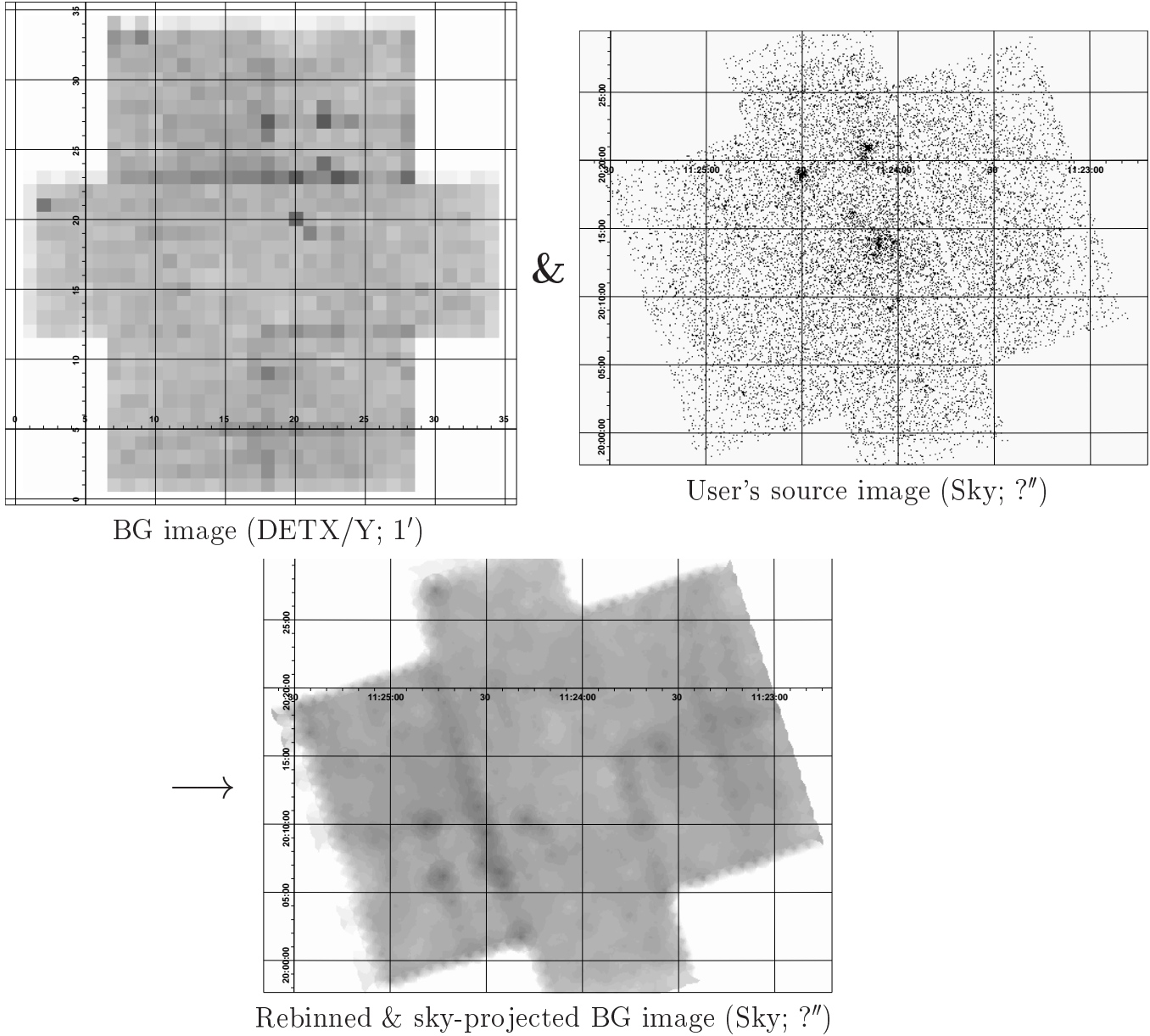}
\caption{Usage of BGrebinimage2SKY to rebin and sky-project a given
low-resolution DETX/Y background map (top left, specifically the M1
thin-filter 0-band particle background map from Fig.\,6) to the
spatial binning and sky position of a user-created high-resolution sky
image (top right). The resultant high-resolution sky background image
is shown at the bottom. The dark areas in the background map are due
mainly to CCD gaps.}
\label{fig1}
\end{figure*}

\section{Event files}

As part of the analysis, for each observation, event files have been
created, in the same manner as Lumb \etal\ (2002), filtered for times
of high background and with all sources removed, as described in
Section\,3. The relevant event files have been merged together for
each instrument/mode/filter combination (Tables 3 \& 4), and eight
separate event files have therefore been created (with nominal
exposure times given as in Table 4), each having an extension
containing a calibrated event list in the same format as produced by
the XMM-SAS. The event files have had sky coordinates assigned to them
for a pointing on the sky of RA=0, Dec=0, PA=0. These event files (or
indeed any) can be reprojected onto any point in the sky via \eg\ {\em
skycast} (see http://www.sr.bham.ac.uk/xmm3/).

It is believed that these event files offer several improvements over
previous versions (\eg\ those of Lumb \etal\ (2002), themselves
improvements on previously-created versions) for several
reasons. Important points are as follows (note that many of these
points apply not only to the event files, but also to the background
maps discussed above);

\begin{itemize}

\item The event files have been created separately for each
combination of instrument, instrument mode and filter. Hence, event
files now exist for medium filter and also for pn full-frame-extended
mode, neither of which had existed previously.

\item As stated in the analysis sections, all data are collected from
source-subtracted and high-background filtered fields with no bright
sources or diffuse features, which could contaminate the `background',
even after source-subtraction.

\item The datasets are longer than previously available event files
(over a million seconds of clean low-background data exists in each of
the thin filter MOS datasets, for example). This improves
signal-to-noise.

\item The secondary extensions of the files have been removed, and all
headers cleaned and corrected. This results in far smaller and more
manageable files. It is possible to treat the event files as normal
SAS event files; images, spectra and lightcurves can be created via
evselect, and one can run ftools on the files.

\item The event files combine a large number (see Table 4) of
observations (rather than a few, long observations), such that the
holes left by the source-subtraction are of less importance, and are
heavily diluted by data from other observations. Due partly to the
different exposure times of the individual observations, no single 
flux limit exists above which sources have been removed. However, flux
histograms of the detected and thereafter removed sources do show
quite a sharp cutoff at $\approx 1\times10^{-14}$\,erg cm$^{-2}$
s$^{-1}$) (see Fig.\,12).

\item The exposure times are believed to be accurate, and incorporate
the effects of losses due to deadtime. For each event file, 4\arcs\
resolution exposure maps (in DETX/DETY coordinates as described in
Table~2) are also available from the web URL
\verb£http://www.sr.bham.ac.uk/xmm3/£. These are useful in evaluating
the degree of the source and bad pixel removal, and calculating the
exact value of the exposure of the background event files in a
particular region of the detector. They are named in a similar manner
to the event files (\eg\ \verb£expmap_ft0000_M1.fits£ is a MOS1
full-frame, thin filter exposure map). The eight exposure maps are
shown in Fig.13.

\end{itemize}

\begin{figure*}
\vspace{10cm}
\includegraphics{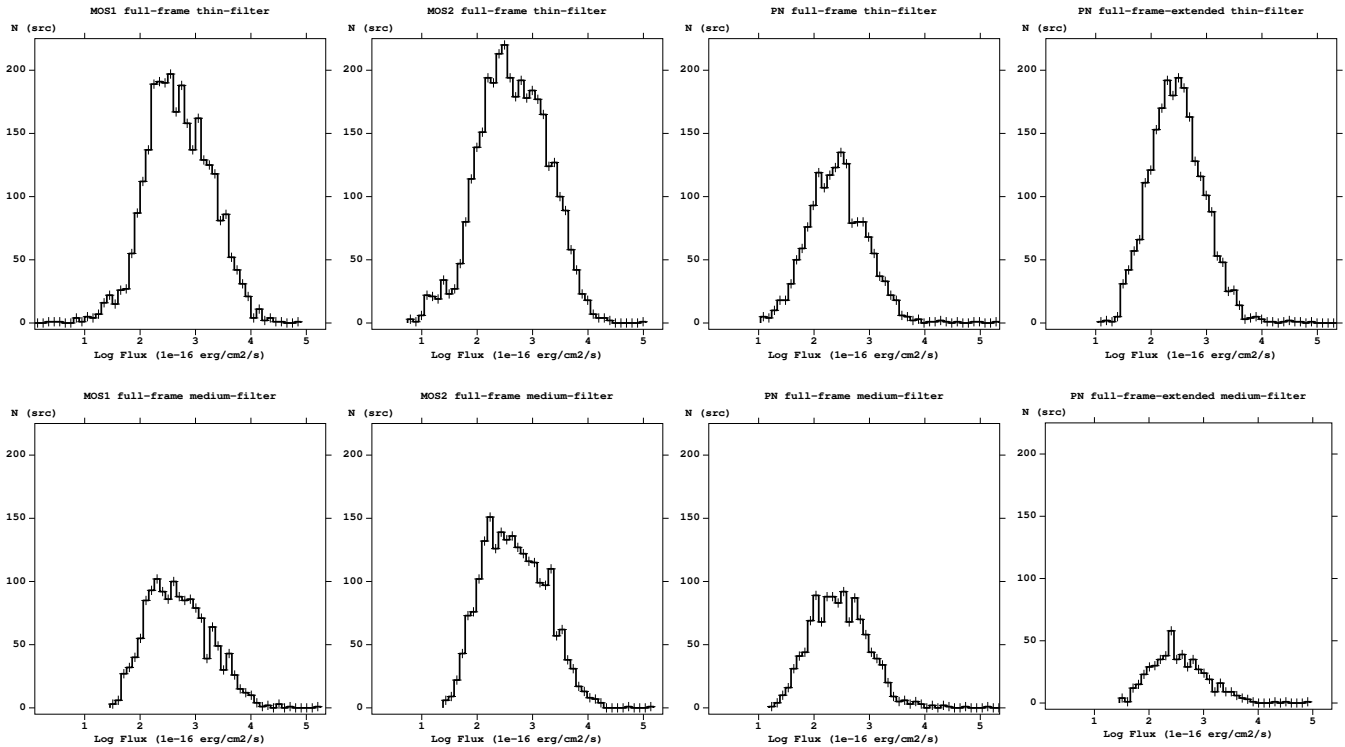}
\caption{Flux histograms of the detected and thereafter removed sources 
  for each of the instrument/mode/filter combinations. (Top) Thin filter:
  MOS1, MOS2 and pn full-frame, pn full-frame extended. (Bottom) the same
  but for medium filter. All plots are to the same scale.}
\label{fig4} 
\end{figure*} 

\begin{figure*}
\vspace{9.5cm}
\includegraphics{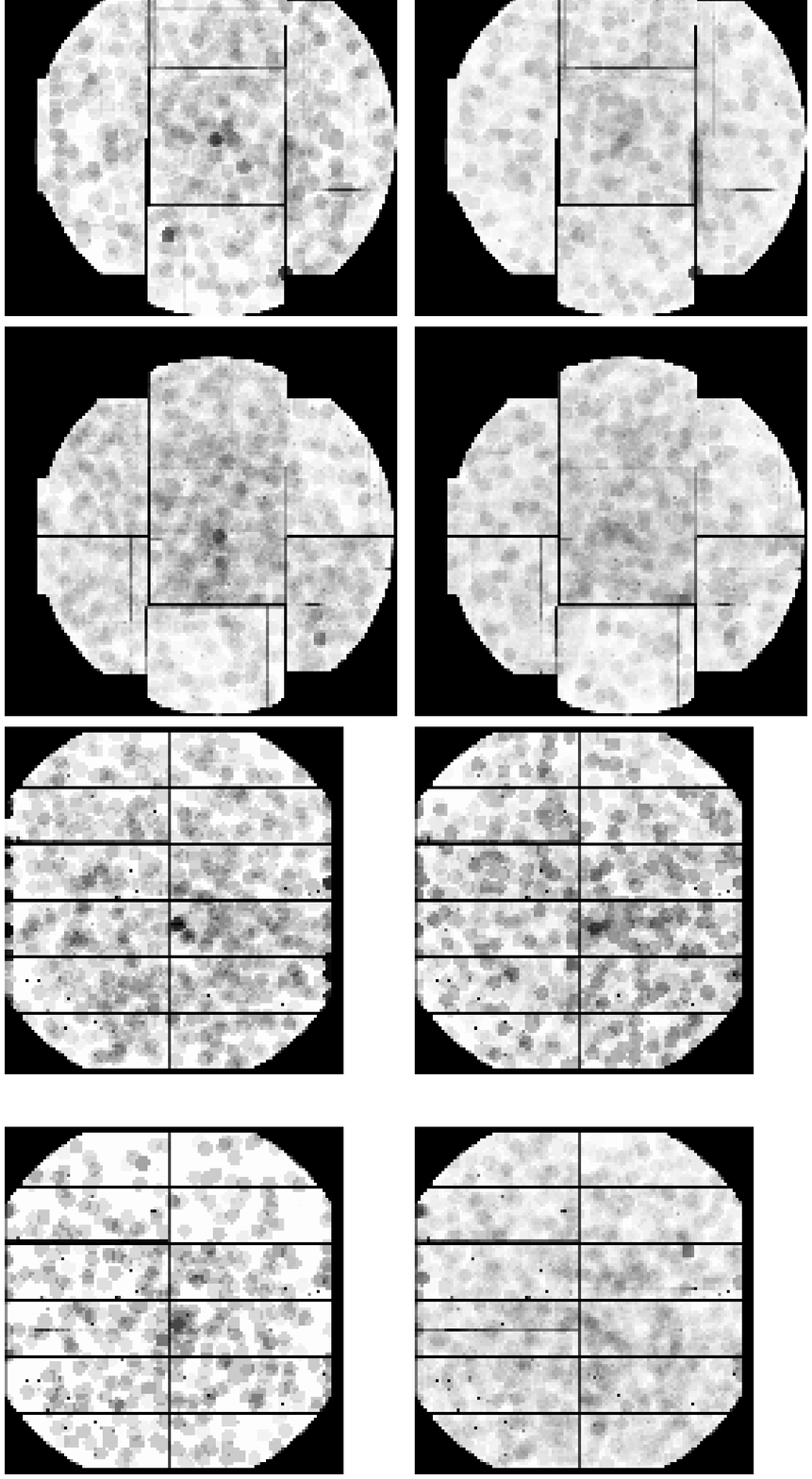}
\caption{4\arcs\ resolution DETX/DETY (see Table 2) exposure maps for
the background event files in each of the instrument/mode/filter
combinations. (Top) Thin filter: MOS1, MOS2 and pn full-frame, pn
full-frame extended. (Bottom) the same but for medium filter.}
\label{fig4} 
\end{figure*} 
\subsection{Event files: usage and caveats}

The eight event files are available from
\verb£http://www.sr.bham.ac.uk/xmm3/£ and are named in a similar
manner to the background maps.  \verb£E1_ft0000_M1.fits£ is a MOS1
full-frame, thin filter event file.  \verb£PN£ and \verb£M2£ are for
the pn and MOS2 instruments, an \verb£e£ denotes extended full-frame
mode, and an \verb£m£, medium filter.

Before using the event files for background analysis, it is strongly
suggested consulting Lumb \etal\ (2002). Because of the above
improvements, a number of the caveats noted by Lumb \etal\ do not
apply to the present files. Many however, are still valid:-

\begin{itemize}

\item Though the extraction of the background can be done in detector
coordinates, the background event files can be {\em skycast} onto the
sky position of the user's field (see http://www.sr.bham.ac.uk/xmm3/ -
alternatively use the XMMSAS task {\em attcalc}).

\item Remnant low-level flares will remain in the event files. It is
known, for instance, that at low energies, the proton flares turn on
more slowly, but earlier than at higher energies. The user could apply
further, more stringent flare-screening to the event files than that
described in Sect.\,3.2, but note that the lowest level proton fluxes
may be spectrally variable, so that no complete subtraction may be
possible.

\item Although point sources have been removed, examination of images
created from the datasets does reveal fluctuations in intensity. Over
scales of arcminutes appropriate to extended sources, this is not
expected to be a significant problem, and indeed representative of
unresolved background. If a user however, should try to extract
spectra from regions comparable to the PSF scale, care should be
taken, and a manual inspection may be necessary to guard against a
local excess or deficit of counts arising from the point source
extraction procedures. In some of the present cases, it is true that,
as many of the observations were pointed such that the target source
was at the centre of the detector, a deficit in counts (by as much as
50\% or greater) is seen at this position. This can be corrected for
by making use of the exposure maps provided (Fig.\,13).

\item Many defects are seen at the lowest energies (below 0.3\,keV),
and the calibration of the EPIC response at these energies is not so
well understood as it is at higher energies. Care should be taken when
performing analysis at \ltsim 0.25\,keV.

\end{itemize}

\section{Future Developments}

At the time of writing, there has been only limited experience gained
in using the maps, the event files and the related software. Any
feedback on using the datasets would be very useful and is most
welcome (please email any comments to \verb£amr30@star.le.ac.uk£).

It is hoped that further releases of the datasets, created using
larger numbers of pointed observations, and perhaps using further
modes and filters, will be made available in the future. This will be
announced in the usual manner and via the URL
\verb£http://www.sr.bham.ac.uk/xmm3/£.

\begin{acknowledgements}
This paper is partially based on data which are proprietary. These
data were made available by the XMM-Newton project scientist
specifically for this purpose. We wish to thank Monique Arnaud, Jean
Ballet, Laurence Jones, Dave Lumb, Silvano Molendi, Wolfgang Pietsch
\& Steve Snowden for very useful discussions during this work, and
Mike Denby for help in obtaining many of the XMM datasets. We also
thank the referee, Fred Jansen, for useful comments which have
improved the paper. AMR acknowledges the support of PPARC funding.
\end{acknowledgements}

\section*{Appendix: The XMM-Newton EPIC X-ray Background: Additional notes}

There have been various studies of the XMM-Newton EPIC background, and
many of these are described in Marty (2002), Lumb (2002) \& especially
Lumb \etal\ (2002).

A number of datasets have also been analysed and merged together, in a
similar manner to that presented here. Two sets of data are
particularly relevant to the present paper:

{\small \rm (i)} Background datasets have been produced (Lumb 2002, Lumb \etal\
2002) for the three EPIC instruments by source-subtracting and
co-adding a few long observations. These event files can be obtained
from \verb£http://xmm.vilspa.esa.es/ccf/epic/#background£ .

{\small \rm (ii)} Event lists combining several CLOSED observations have been created by
Phillipe Marty, and these have proven very useful in the analysis and
background-subtraction of extended objects (see Sect.3.2). These data
can be obtained from
\verb£ftp://www-station.ias.u-psud.fr/pub/epic/Closed£ 

In addition, several novel methods have been used to analyse very
extended and diffuse X-ray sources, where the extraction of the
background is difficult. Many of these are described in Marty \etal\
2002. Of special interest are the works of Arnaud \etal\ (2001, 2002),
de Luca \& Molendi (2002), Ghizzardi \etal\ (2000), Molendi (2001),
and Pratt \& Arnaud (2002).

\end{document}